# Anomalous magneto-transport in disordered structures: classical edge-state percolation

Walter Schirmacher,[1,2] Benedikt Fuchs,[3] Felix Höfling,[4] and Thomas Franosch[1,*]

[1] Institut für Theoretische Physik, Leopold-Franzens-Universität Innsbruck, Technikerstraße 21a, A-6020 Innsbruck, Austria
[2] Institut für Theoretische Physik, Johannes Gutenberg-Universität Mainz, Staudinger Weg 7, D-55088 Mainz, Germany
[3] Institut für Wissenschaft Komplexer Systeme, Medizinische Universität Wien, Spitalgasse 23, A-1090 Wien, Austria
[4] Max-Planck-Institut für Intelligente Systeme, Heisenbergstraße 3, 70569 Stuttgart, Germany, and
IV. Institut für Theoretische Physik, Universität Stuttgart, Pfaffenwaldring 57, 70569 Stuttgart, Germany

By event-driven molecular dynamics simulations we investigate magneto-transport in a two-dimensional model with randomly distributed scatterers close to the field-induced localization transition. This transition is generated by percolating skipping orbits along the edges of obstacle clusters. The dynamic exponents differ significantly from those of the conventional transport problem on percolating systems, thus establishing a new dynamic universality class. This difference is tentatively attributed to a weak-link scenario, which emerges naturally due to barely overlapping edge trajectories. We make predictions for the frequency-dependent conductivity and discuss implications for active colloidal circle swimmers in a heterogeneous environment.

PACS numbers: 64.60.ah,75.47.-m

Electronic transport in two-dimensional (2D) disordered structures under the influence of a magnetic field exhibits a fascinating wealth of observed phenomena [1–11], including the quantum Hall effect [12–15]. The field has gained new momentum by recent investigations on disordered graphene [16–18] and other topological insulators [19, 20]. For many of these phenomena classical magneto-transport constitutes the basis of a semi-classical description [13, 21–23]. For instance, the edge states in quantum Hall systems are the quantum analogue of "skipping orbits", trajectories formed by circular arcs bouncing along the edges of a mesoscopic structure.

One of the widely investigated classical models for transport in disordered systems is the Lorentz model [24, 25]. In this model a particle is specularly scattered by circular or spherical obstacles, which may overlap and are distributed randomly according to a Poisson process. Transport within the Lorentz model and the associated percolation transition have been studied extensively in the past [26–33]. The Lorentz model has been considered also in the presence of an applied magnetic field [1–6, 34–38], i.e. the linear paths between successive specular scattering events are replaced by circular arcs. It turned out that in the presence of the field a description of the transport in terms of the Boltzmann equation is no longer appropriate because of the breakdown of both ergodicity [23, 39] and the Markov property of the scattering sequence [3–5]. The resulting strongly correlated kinetics exhibits a rich scenario of anomalous magneto-resistance [1–4, 34–37]. Let us note that the circular motion is not only realized by electrons in a magnetic field, but also by active particles subject to asymmetric driving [40–45]. Such particles in the presence of randomly distributed obstacles [46, 47] provide a colloidal analogue of the Lorentz model with magnetic field.

An interesting feature of this Lorentz model is the existence of a magnetic-field-induced localization, which is of percolative character [1, 2, 4, 6]. Via the classical–quantum correspondence this transition is also relevant for the quantum localization in a magnetic field. Very recently, a theoretical investigation of spin-Hall topological insulators with random circular obstacles has shown that, because of edge-state percolation, a similar insulator–conductor transition emerges [48]. For the magnetic transition a relation for the field-dependent critical density has been derived by Kuzmany and Spohn [6]. A detailed numerical investigation of this transition and the associated critical transport has remained a challenge so far.

Here, we present results of large-scale molecular dynamics (MD) simulations for the Lorentz model with circular motion. We focus on the field-induced localization transition and investigate the nature of the trajectories leading to critical slowing down and anomalous diffusion. In particular we determine the static and dynamic critical exponents both for the conventional and the magnetic transition and argue for a new universality class of the latter. We shall present a heuristic argument for the suppression of transport with respect to the standard transport scenario on percolating systems.

Our setup describes a two-dimensional gas of classical, independent carriers of charge $q$ and mass $m$ in a random array of overlapping hard-disk obstacles of radius $\sigma$ in the presence of a perpendicular, uniform magnetic field $B$. The particles move with constant velocity $v$, and the trajectories become *skipping orbits* [49] consisting of circular arcs with cyclotron radius $R = mv/qB$ connected by specular scattering events (Fig. 1). We employ event-driven MD simulations similar to the field-free case [28, 33, 50]. Furthermore we consider uniformly distributed obstacles of number density $n$, leaving as control parameters the dimensionless density $n^* = n\sigma^2$ and the dimensionless ratio $R/\sigma$. At densities $n^* > n_c^* = 0.359081\ldots$ [29, 31, 51] the accessible void space consists only of disconnected pockets, prohibiting long-range transport. At lower densities an infinite void-space cluster emerges, and a finite diffusivity $D$ arises if the particles can permeate the entire system via skipping orbits, see Fig. 1. At low densities the skipping-orbit motion occurs only around isolated obstacle clusters, which corresponds to a magnetic-field-induced insulating phase (topological insulator). Whereas deep in the conductive phase the trajectories are dominated by many scattering events similar to the field-free case, close to the transition the motion is characterized by regular skipping orbits jumping occasionally between isolated obstacle clusters.

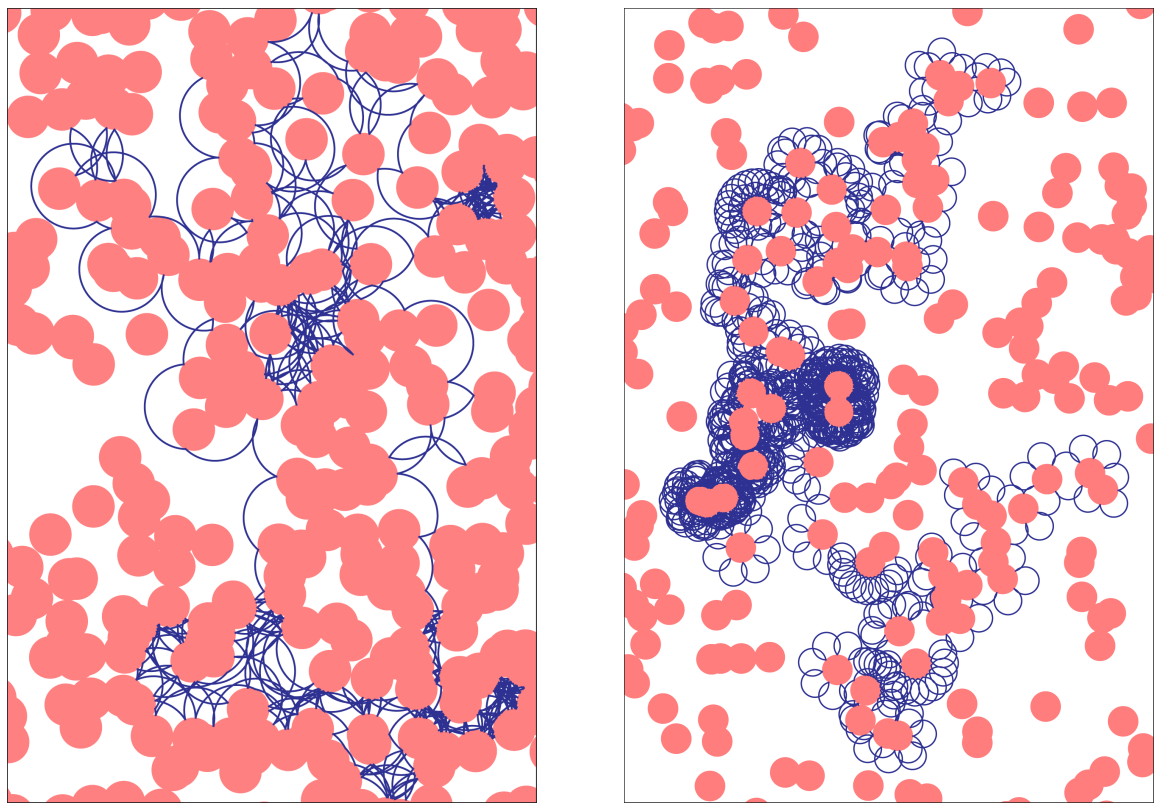

FIG. 1. Typical trajectories for classical magneto-transport. Left: conductive phase well above the magnetic transition ($R/\sigma$ = 2.0, $n^*$ = 0.3); right: almost at the transition ($R/\sigma$ = 0.9, $n^*$ = 0.1), see the red dots in Fig. 4.

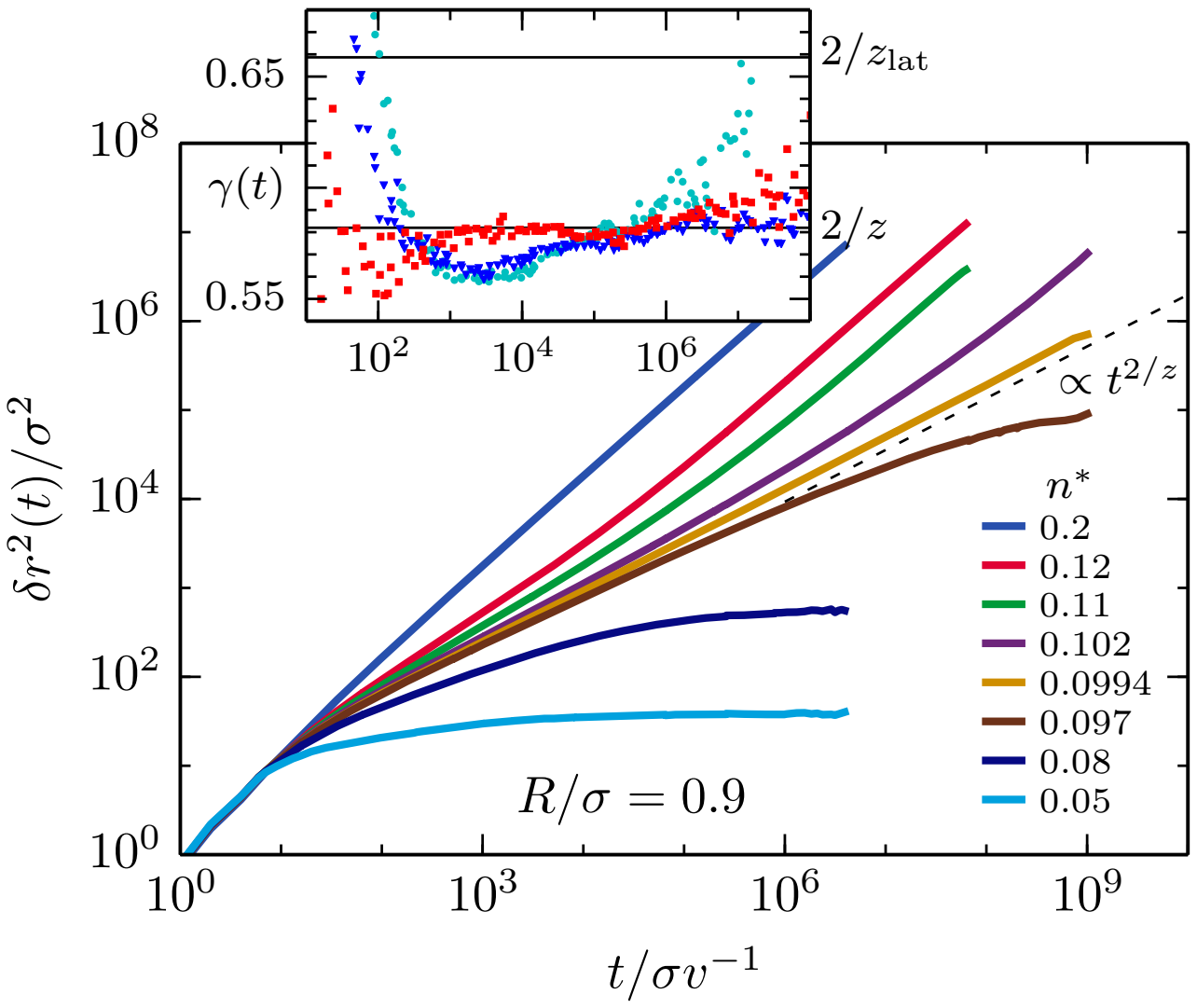

FIG. 2. Mean-square displacements $\delta r^2(t)$ for $R/\sigma = 0.9$. The density $n^*$ decreases from top to bottom. The dashed line indicates the critical asymptote $\propto t^{2/z}$ with $z = 3.43$. Inset: The local exponent $\gamma(t) := \mathrm{d}\log\big(\delta r^2(t)\big)/\mathrm{d}\log(t)$ as a function of time at the magnetic transition, $n^* = n^*_m(R)$, for three cyclotron radii: $R/\sigma$ = 0.5 (cyan circles), 0.9 (blue triangles), and 2.0 (red squares). The two horizontal lines mark the subdiffusion exponent for the universality classes of lattice percolation ($2/z_{\mathrm{lat}}$) and of the magnetic localization transition ($2/z$), respectively.

We have determined the mean-square displacement (MSD) $\delta r^2(t) := \big\langle[\boldsymbol{R}(t) - \boldsymbol{R}(0)]^2\big\rangle$ as averages over time, tracer ensemble, and disorder. We have used large system sizes of $10^4\sigma$ with periodic boundaries. As an example we show data for $R/\sigma = 0.9$ in Fig. 2. For moderate densities ($n^* \gtrsim 0.1$) the MSD grows linearly in time $\delta r^2(t \to \infty) \simeq 4Dt$ for long times, where $D$ is the diffusion coefficient. Decreasing the density $n^*$ this linear regime is delayed to longer and longer times, until eventually at a critical density $n^*_m = n^*_m(R)$ the long-time behavior becomes *sub*-diffusive, $\delta r^2(t \to \infty) \sim t^\gamma$, with an observed exponent of $\gamma = 0.581 \pm 0.005$ [52]. The inset of Fig. 2 shows a rectification by means of the local exponent $\gamma(t) := \mathrm{d}\log\big(\delta r^2(t)\big)/\mathrm{d}\log(t)$ as a function of time, corroborating the long-time asymptotics. Such fractional (or anomalous) diffusion is found widely in complex systems, but the physical origins are often difficult to pin down [54–56]; here it emerges naturally from a critical phenomenon. The anomalous exponent is related to the dynamic critical exponent $z$ via $\gamma = 2/z$, which gives $z = 3.44 \pm 0.03$. This value is incompatible with the known dynamic exponent $z_{\mathrm{lat}} = 3.036 \pm 0.001$ (corresponding to $\gamma_{\mathrm{lat}} = 0.658$, see the inset of Fig. 2) for two-dimensional random walkers on percolating lattices [30, 57], valid also for the field-free localization transition [29, 31, 58, 59]. Although the values of $z$ and $z_{\mathrm{lat}}$ differ only by 10%, the data cannot be described in a satisfactory way [52] with an assumed exponent $z_{\mathrm{lat}}$, even if corrections to scaling [30] are included. Decreasing the density below $n^*_m$ the MSD converges at long times $\delta r^2(t \to \infty) = \ell^2$, which defines $\ell$ as the localization length.

The data close to the magnetic transition suggest a dynamic scaling scenario similar to the localization transition for transport on percolating clusters [31]. Here the localization length $\ell$ is identified with the mean-cluster size diverging as $\ell \sim |\varepsilon|^{-(\nu-\beta/2)}$ upon approaching the transition, where $\varepsilon = (n^* - n^*_m)/n^*_m$ is the dimensionless separation parameter. The exponent $\nu$ quantifies the divergence of the largest finite cluster of linear extension $\xi \sim |\varepsilon|^{-\nu}$ (correlation length), and $\beta$ measures the weight of the infinite cluster $\sim \varepsilon^\beta$ for $n^* > n^*_m$ (order parameter). In 2D, the values for standard percolation are known exactly: $\nu = 4/3, \beta = 5/36$ [57]. We verified by means of a rectification plot that, indeed, our data for $\ell$ are compatible with these values (Fig. 3a).

The MSD is expected to obey dynamic scaling, $\delta r^2(t) = t^{2/z}\delta\hat{r}^2_\pm(\hat{t})$, with scaling functions $\delta\hat{r}^2_+(\cdot)$ and $\delta\hat{r}^2_-(\cdot)$ for the conducting and the insulating side, respectively. Here time enters only in a rescaled way $\hat{t} = t/t_x$ with a crossover time $t_x \sim \ell^z$. The scaling functions become constant for small argument, $\delta\hat{r}^2_\pm(\hat{t} \ll 1) = const$ (critical regime), whereas for large argument they approach $\delta\hat{r}^2_+(\hat{t} \gg 1) \sim \hat{t}^{1-2/z}$ on the conducting side and $\delta\hat{r}^2_-(\hat{t} \gg 1) \sim \hat{t}^{-2/z}$ on the insulating side. Hence one infers that the diffusion coefficient vanishes as $D \sim \varepsilon^\mu$ with a *magnetic* conductivity exponent $\mu = (z - 2)(\nu - \beta/2) = 1.82 \pm 0.04$. This exponent differs significantly from the corresponding standard value for random resistor networks on lattices, $\mu_{\mathrm{lat}} = 1.310$ [27, 30].

We have measured diffusivities $D = D(n, R)$ throughout the phase diagram, see Fig. 4. In particular, we have verified the magnetic transition scenario also for two other values for the cyclotron radius and found similar results for the conductivity exponent at $R/\sigma = 0.5$ and $R/\sigma = 2.0$ [52][60]. In order to judge, whether the slight variation of these values with $R$ is significant or just due to asymptotic corrections and the limited accuracy, we made a rectification plot (Fig. 3b) with the *same* value $\mu = 1.82$ for all values of $R$ and obtain satisfactory rectifications. We conclude that, within our accuracy, the dynamic critical exponents of the magnetic transition have values $\mu = 1.82 \pm 0.08$ and $z = 3.44 \pm 0.06$ [52] and do not depend on the field parameter $1/R$.

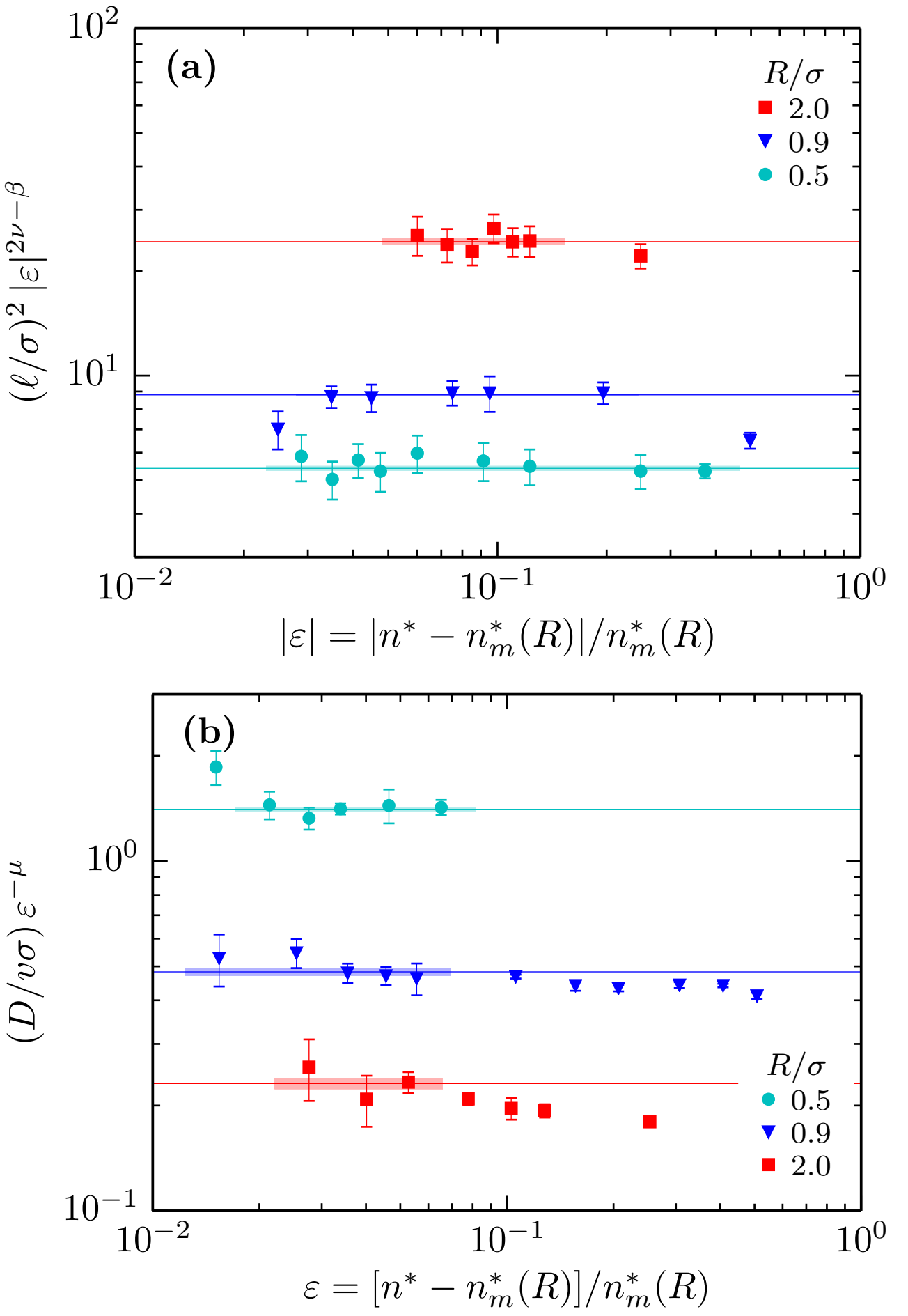

FIG. 3. a) Rectification plot of the squared localization length $(\ell/\sigma)^2|\varepsilon|^{2\nu-\beta}$ vs. $|\varepsilon| = |n^* - n^*_m(R)|/n^*_m(R)$ in the insulating phase for different cyclotron radii $R$. Error bars combine statistical errors and uncertainties in reading off $\ell^2$ from the long-time asymptotes of $\delta r^2(t)$. Thin solid lines indicate weighted averages of the data points and shaded bars the uncertainty of the respective mean value; only data points covered by the bars were taken into account. b) Rectification plot of the diffusion constant $(D/v\sigma)\varepsilon^{-\mu}$ vs. $\varepsilon$ in the conducting phase. Data for different $R$ are shifted downwards by factors of 2. For all the three values of $R$ the same conductivity exponent $\mu = 1.82$ was used.

As mentioned in the introduction, a relation between the critical density $n^*_m$ of the magnetic localization transition and the applied field $B$ has been suggested by Kuzmany and Spohn [6]. They argued in terms of the cyclotron radius $R = mv/qB$ that one should consider the percolation of disks of effective

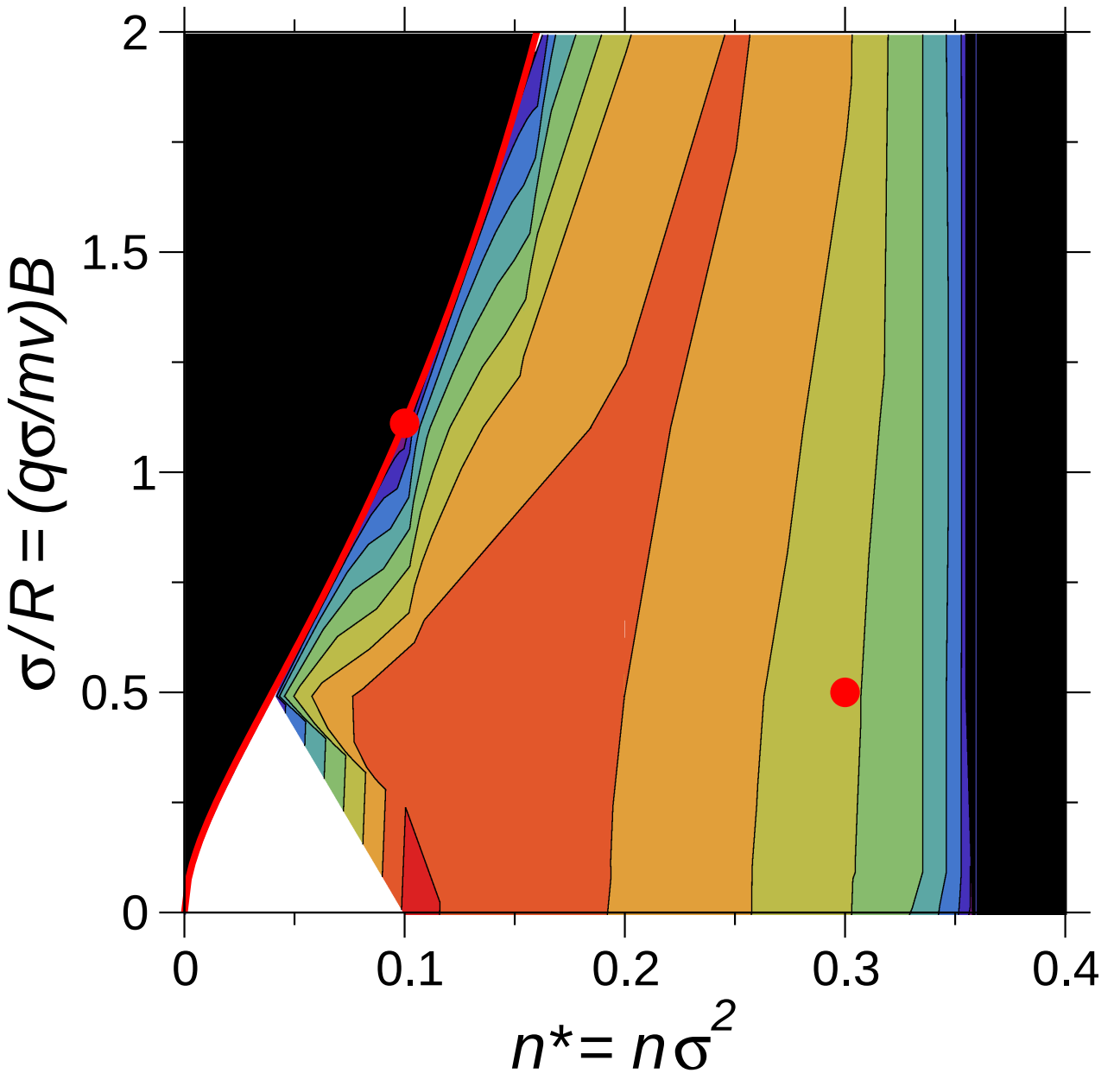

FIG. 4. Phase diagram: magnetic field $\sigma/R \propto B$ vs. density $n^*$. The isodiffusivity contours are spaced logarithmically and increase from the phase boundaries towards the inner region. The red line corresponds to the magnetic transition according to Kuzmany and Spohn [6], $n^*_m(R) = n^*_c\sigma^2/(\sigma+R)^2$. The big red dots indicate the parameters for the trajectories of Fig. 1.

radius $R + \sigma$. This argument leads to a critical density given by

$$n^*_m(R)(\sigma + R)^2/\sigma^2 = n^*_c = 0.359081\ldots, \qquad (1)$$

this line of magnetic transitions is included in Fig. 4. We have verified that the critical magnetic density $n^*_m$ observed in our simulations coincides with this prediction to at least two significant digits at the three values of $R$ explored, which suggests that Eq. (1) is an exact relation.

The phase diagram displays a second transition line which is independent of the magnetic field and occurs at the percolation density $n^*_c$. We have verified that the values of the scaling exponents characterizing this localization transition are the same as in the field-free case, irrespective of the magnitude of the magnetic field.

Why do the values for $\mu$ and $z$ differ from their conventional ones? It is well known that the dynamic exponents for transport on percolating systems may differ from those of the standard random resistor network due to the presence of weak links. It has been demonstrated by means of mapping the traditional nodes-links-blobs model of percolation clusters to conductance networks [61] that if the distribution of conductances $\Gamma$ along weak links obeys a singular statistics, $\varrho(\Gamma) \propto \Gamma^{-a}$ with $0 < a < 1$, there exists an exponent relation

$$\mu = \max\left[\mu_{\text{lat}}, (d-2)\nu + 1/(1-a)\right], \qquad (2)$$

where $d$ is the dimension of the embedding space. This re-lation has been shown to be exact to all orders in a renormalization group $\varepsilon$-expansion [62]. The conductances can also be interpreted as transition probabilities across the weak links. Machta and Moore [63] have shown that weak links emerge in the (field-free) Lorentz model and lead to a value of $a = (d-2)/(d-1)$. Accordingly, in $d = 3$ weak links dominate the transport while in $d = 2$ the standard random-resistor exponents are valid.

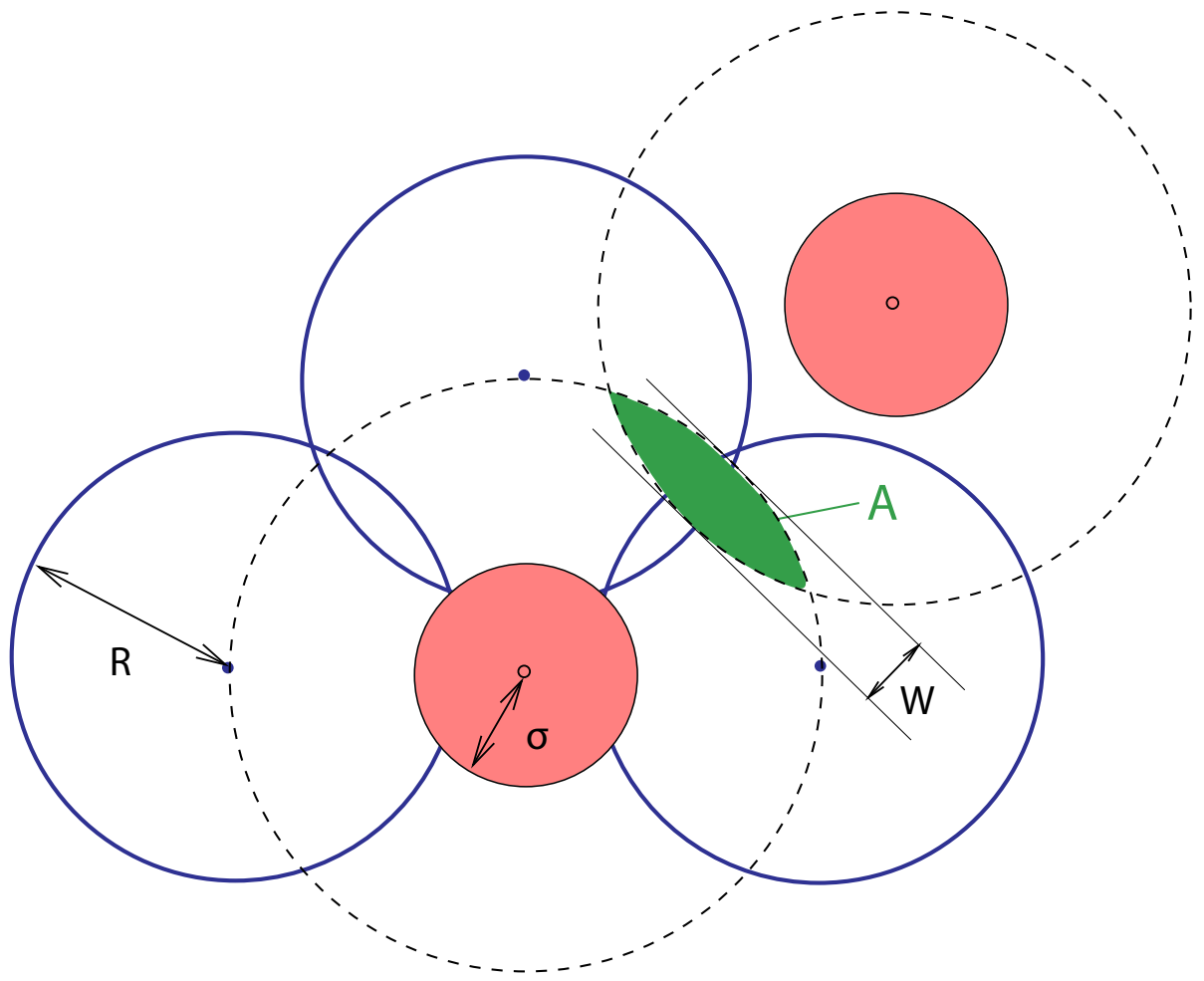

FIG. 5. Visualization of the weak-link scenario of the field-induced percolation transition. Filled (apricot) circular regions are obstacles, the continuous (blue) arcs form a skipping orbit, and dashed circles are loci of skipping-orbit centers. The (green) overlap area of center loci scales as $A \sim W^{3/2}$ for vanishing width $W$.

If the weak-link scenario is responsible for the modified value $\mu$, a certain value of $a = 1 - 1/\mu$ for the distribution of weak links has to be rationalized. As noted above, near the magnetic transition the centers of the skipping orbits move along the perimeter of a percolation cluster formed by circles of radius $\sigma + R$. This cluster is generated by isolated obstacle clusters, see the right panel of Fig. 1. A weak-link scenario can be identified considering the transitions of the trajectories between these clusters. The weak links correspond to configurations at which the circles formed by the centers of the skipping orbits have a very small overlap, see Fig. 5. As the two Cartesian components of the centers of the skipping orbits are canonically conjugate to each other [14], their possible values inside the overlap area $A$ represent the phase space for the transition. Thus the transition probability across the weak link is proportional to this overlap region $A$. Let us speculate how the decrease of the overlap region $A$ induces a suppression of transport, i.e. an increase of the conductivity exponent $\mu$. Following Machta and Moore [63], the probability density $P(W)$ for the width $W$ of the overlap of randomly distributed disks (Fig. 5) approaches a constant $P(W \to 0) \neq 0$ in the limit $W \to 0$. By geometric considerations one can work out that $A \sim W^{3/2}$ for $W \to 0$. Thus the probability density $\varrho(\Gamma)$ of the transition rates $\Gamma \propto A$ satisfies $\varrho(\Gamma) = P(W)\,\mathrm{d}W/\mathrm{d}\Gamma \sim P(W \to 0)\,\mathrm{d}W/\mathrm{d}A \sim A^{-1/3} \sim \Gamma^{-1/3}$ and thus $a = 1/3$. By virtue of the hyperscaling relation, Eq. (2), this leads to a value of $\mu = 3/2$ which is closer to the observed value $\mu \approx 1.82$ than the standard conductivity exponent $\mu_{\mathrm{lat}} = 1.310$.

In conclusion we have studied for the first time dynamic critical behavior of the low-density, magnetic-field-induced localization transition in the Lorentz model, using high-quality data obtained by state-of-the-art event-driven MD simulations. We have identified that this transition comprises a new universality class of dynamic percolation in that the dynamic exponents are different from their counterparts in conventional percolation problems.

We were able to corroborate a weak-link scenario for transitions between the barely overlapping edge states. Remarkably, weak links are relevant for the magnetic 2D Lorentz model with magnetic field—in contrast to the usual case (without field) where weak links are only of importance in $d = 3$ [28, 30, 31, 63]. We mention also that near the magnetic transition the path described by the centers of the skipping orbits constitutes a disordered topological insulator, as this directed path runs along the perimeter of an effective percolating cluster of disks.

Our findings have direct implications for the complex frequency-dependent conductivity $\Sigma(\omega)$, measurable in a conventional transport set-up for magneto-resistance of a 2D electron gas [64]. By the Einstein–Kubo relation $\Sigma(\omega) \propto Z(\omega)$ [65], where $Z(\omega)$ is the Fourier–Laplace transform of the velocity autocorrelation function, the subdiffusive motion directly at the transition translates to an anomalous power-law dispersion $\Sigma(\omega \to 0) \sim \omega^{1-2/z}$ [33, 66], with a rich cross-over scenario as $n^* \downarrow n^*_m$ (see Ref. [52] and Fig. S3 therein).

Finally, we point out that the field-induced percolation transition, which is entirely of geometric origin, may also be realized experimentally by colloidal circle swimmers [40–45] in a heterogeneous environment [46, 47]. If in our model the condition of specular reflection at the obstacles is replaced by an appropriate distribution of random reflections [47], the maximum distance between the center of an orbit and an obstacle is still $\sigma + R$, and the condition for a *curvature-induced* percolation transition is also given by Eq. (1). It will be worth wile to study this transition in the future.

This work has been supported by the Deutsche Forschungsgemeinschaft DFG via the Research Unit FOR1394 "Nonlinear Response to Probe Vitrification".

* Corresponding author: thomas.franosch@uibk.ac.at

# Supplement to
# "Anomalous magneto-transport in disordered structures: classical edge-state percolation"

Walter Schirmacher, Benedikt Fuchs, Felix Höfling, and Thomas Franosch

## I. ANALYSIS OF THE CRITICAL DYNAMICS

In this section, it will be shown that for the magnetic localization transition discussed in the main paper the dynamic exponents $\mu$ and $z$ deviate significantly from their lattice values. Data sets for three different field strengths, corresponding to cyclotron radii $R/\sigma = 0.5,\ 0.9,\ 2$ are analyzed. For each value of the cyclotron radius $R$, a series of reduced obstacle densities $n^*$ near the magnetic transition line $n_m^*(R) = n_c^*(1 + R/\sigma)^{-2}$ is considered, anticipating a prediction by Kuzmany and Spohn [1].

The mean-square displacements (MSDs) contain information on the critical dynamics from two perspectives. First for near-critical obstacle densities, the long-time asymptotes of the MSD yield the diffusion constant $D$ and the localization length $\ell$, respectively, as functions of the distance $\varepsilon = (n^* - n_m^*)/n_m^*$ to the transition. Both quantities display critical singularities,

$$D \sim \varepsilon^{\mu} \qquad (\varepsilon \downarrow 0)\,, \tag{S1}$$

$$\ell \sim |\varepsilon|^{-\nu+\beta/2} \qquad (\varepsilon \uparrow 0)\,, \tag{S2}$$

from which the conductivity exponent $\mu$ and the combination $\nu - \beta/2$ can be inferred. Second, the MSD is asymptotically subdiffusive at the transition,

$$\delta r^2(t \to \infty; n^* = n_m^*) \sim t^{2/z}\,, \tag{S3}$$

which allows one to determine the dynamic exponent $z$. Finally, the exponent relation

$$z = 2 + \mu/(\nu - \beta/2) \tag{S4}$$

serves as a consistency check.

Anticipating universality, precise values for the exponents are known from random walkers on percolation lattices, mappings to random resistor networks, and conformal field theory. In two dimensions, $\nu = 4/3$, $\beta = 5/36$ [2], and $\mu_{\text{lat}} = 1.310(1)$, which implies $z_{\text{lat}} = 3.036(1)$ [3, 4]; the numbers in parentheses indicate the uncertainty in the last digit. A sensitive test whether the expected universality class describes a given data set is achieved by so-called *rectification plots*. For example, the product $D\varepsilon^{-\mu}$ with given exponent $\mu$ should approach a constant value for $\varepsilon \downarrow 0$.

If the value of the exponent is not known *a priori* (e.g., from theoretical considerations), data interpretation in terms of critical power laws is a delicate issue for two reasons. First, the prediction of a power-law is only asymptotic: it should hold better and better for longer and longer times, such that power-law-like corrections slowly fade out. Often data appear to follow straight lines on double-logarithmic scales and it is tempting to ignore systematic drifts in the slope, thereby merely assigning effective exponents. Second, the quality of the data becomes worse as longer times are probed, both from statistical fluctuations (individual trajectories) but also systematic variations (finite size, sample-to-sample fluctuations, uncertainties in the critical point). As a consequence of these effects, the error bars obtained blindly from regression algorithms typically underestimate the actual error significantly.

### A. Localization lengths

Figure 3a of the main paper establishes that the product $\ell^2|\varepsilon|^{2\nu-\beta}$ is independent of $\varepsilon$ for the investigated range of densities. Deviations are merely found for the largest values of $|\varepsilon|$, as expected. At $R/\sigma = 0.9$, the data point at the smallest $|\varepsilon|$ was discarded as an outlier, possibly because the corresponding data for $\delta r^2(t)$ do not reach sufficiently far into the asymptotic regime. We conclude with a high level of confidence that the localization length $\ell$ diverges at the transition with the exponent $\nu - \beta/2$ attaining its value on lattices. In passing, the data also substantiate the Kuzmany and Spohn [1] prediction for $n_m^*(R)$.

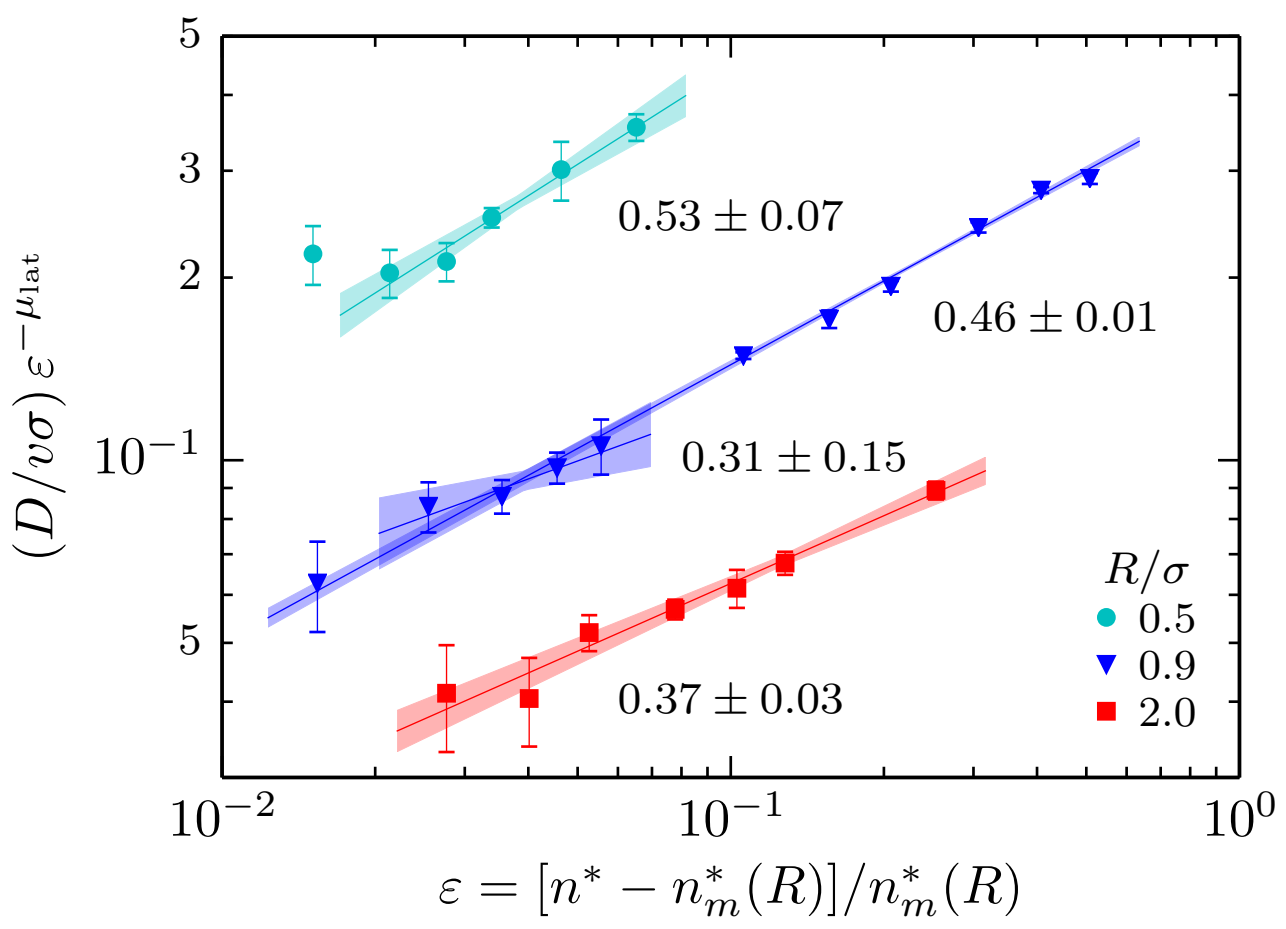

Figure S1. Rectification plot of the diffusion constant $D$ for obstacle densities above the magnetic transition, using the lattice value $\mu_{\text{lat}}$ for the conductivity exponent. Deviations from the expected exponent are tested by a weighted power-law regression. The results and their uncertainties are shown by straight lines and shaded areas, respectively. Numbers next to each line give the obtained slope $\Delta\mu$. Data for different $R$ are shifted downwards by factors of 2.

### B. Diffusion constants

For the diffusion constants, let us first assume that the magnetic transition falls into the same dynamic universality class as standard transport on percolation lattices. Rectification with the lattice exponent $\mu_{\mathrm{lat}}$, however, does not show convergence as $\varepsilon \downarrow 0$ (Fig. S1). Rather, the data for $D\varepsilon^{-\mu_{\mathrm{lat}}}$ on double-logarithmic scales are well described by straight lines with slope $\Delta\mu$, indicating a power law $D \sim \varepsilon^{-(\mu_{\mathrm{lat}}+\Delta\mu)}$ over almost the whole density range investigated. The data provide strong evidence that the value $\mu_{\mathrm{lat}}$ does not describe the critical dynamics of the magnetic-field-induced localization transition. Rather, the exponent adopts a new value $\mu = \mu_{\mathrm{lat}} + \Delta\mu$. At $R/\sigma = 0.9$, the data follow $D \sim \varepsilon^{\mu}$ with this value of the exponent over almost two decades in $\varepsilon$.

For each cyclotron radius $R$, we have performed a "power-law regression" of $D\varepsilon^{-\mu_{\mathrm{lat}}}$ as function of $\varepsilon$ with weights given by the uncertainty of each data point, combining statistical and read-off errors. Details of the fitting procedure are given in Appendix A, the obtained estimates and their uncertainties for the slope $\Delta\mu$ are shown in Fig. S1. Increasing the magnetic-field strength from $R/\sigma = 2$ to 0.5, the deviation appears to increase from $\Delta\mu \approx 0.37 \pm 0.03$ to $0.53 \pm 0.07$. Taking into account both statistical errors and possible systematic errors due to insufficient asymptotics, in all cases the obtained slopes $\Delta\mu$ are definitely non-zero and attain values much larger than the uncertainty of the data. (A worst case scenario for $R/\sigma = 0.9$ has been included in the figure, using only 4 data points for the fit.) This means that the field-induced vanishing of the diffusion constant cannot be described by the standard universality class.

### C. Mean-square displacements

The second approach to the critical dynamics is based on the subdiffusive growth of the MSD precisely at the localization transition, $\delta r^2(t) \sim t^{2/z}$. Since it is much easier to generate data for large intervals in time rather than $\varepsilon$, this approach typically yields more accurate results. A rectification test using the value $z_{\mathrm{lat}}$ for the dynamic exponent fails clearly (Fig. S2). On double-logarithmic scales, the data follow straight lines over up to 7 orders of magnitude in time with negative slopes $-\Delta\gamma$ deviating significantly from zero (middle part of Table S1).

| $R/\sigma$ | 0.5 | 0.9 | 2.0 |
|---|---|---|---|
| $n^*$ | 0.1600 | 0.0994 | 0.0399 |
| fit range for $t$ | $(10^5, 10^7)$ | $(10^5, 10^8)$ | $(10^3, 10^7)$ |
| $\Delta\gamma$ | 0.066(3) | 0.076(2) | 0.077(1) |
| $\Delta z$ | 0.341(9) | 0.395(5) | 0.404(4) |
| $\Delta\mu$ | 0.43(1) | 0.499(7) | 0.510(4) |
| $\gamma$ | 0.59(2) | 0.581(5) | 0.582(5) |
| $z$ | 3.37(8) | 3.44(3) | 3.44(3) |
| $\mu$ | 1.7(1) | 1.82(4) | 1.81(4) |

Table S1. Results from the analysis of the rectified MSDs (middle part, Fig. S2) and the local exponents $\gamma(t)$ (bottom part, inset of Fig. S2). The slopes $-\Delta\gamma$ were obtained from power-law regression fits. The remaining lines were calculated from $\Delta z := z - z_{\mathrm{lat}}$ with $2/z := 2/z_{\mathrm{lat}} - \Delta\gamma$ and the exponent relation Eq. (S4). The numbers in parentheses specify the uncertainty in the last digit as obtained from the regression method.

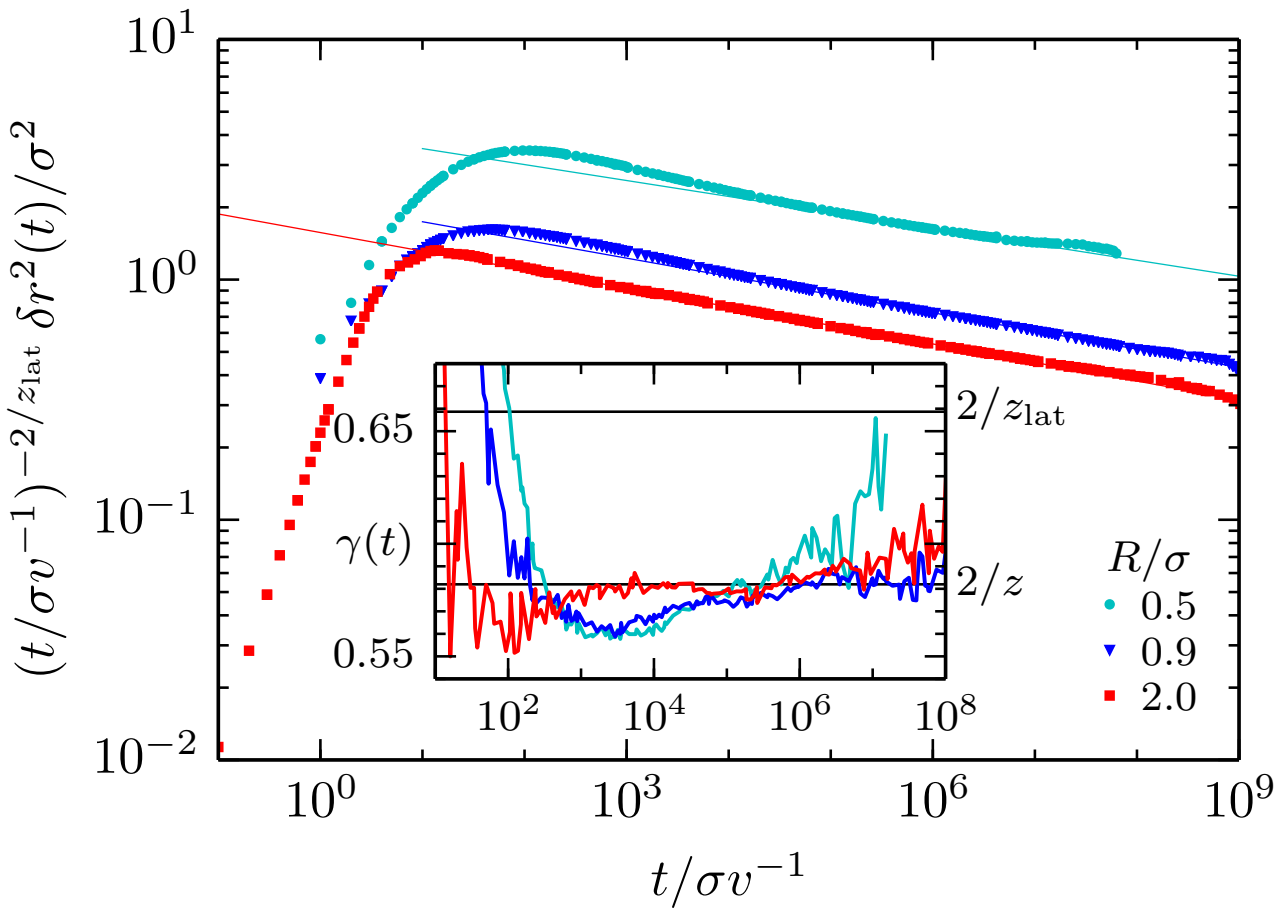

Figure S2. Rectification plot of the mean-square displacement $\delta r^2(t)$ based on the lattice value $z_{\mathrm{lat}}$ for the dynamic exponent. On double-logarithmic scales, the obtained straight lines for times $t \gg 10^3\sigma v^{-1}$ suggest subdiffusion but with a different dynamic exponent $z$ (see Table S1). Data for different $R$ are shifted downwards by factors of 2. Inset: The local exponent $\gamma(t)$, defined in Eq. (S5), does not converge to $2/z_{\mathrm{lat}}$, rather a new value $2/z$ is adopted.

An alternative analysis of the MSD is based on the local exponent

$$\gamma(t) := \mathrm{d}\log\bigl(\delta r^2(t)\bigr)/\mathrm{d}\log(t). \qquad \text{(S5)}$$

Numerically, we found that the simple scheme of finite differences is sufficient to obtain reasonable results for $\gamma(t)$, see inset of Fig. S2 and bottom part of Table S1. (Note that we have evaluated the correlation functions at lag times which are approximately uniformly distributed on the logarithmic scale.) For all three values of $R$, the local exponent falls clearly below $2/z_{\mathrm{lat}}$ near $t \approx 10^2\sigma v^{-1}$ and remains at low values for longer times. In particular, there is no tendency to approach $2/z_{\mathrm{lat}}$ at long times. At weak magnetic field ($R/\sigma = 2$), the local exponent converges quickly to its apparent long-time limit, $\gamma(t \to \infty) = 2/z \approx 0.582(5)$, which is more than 10% smaller than the value $2/z_{\mathrm{lat}} \approx 0.659$ on lattices. The data for $\gamma(t)$ at $R/\sigma = 0.9$ approach the *same* limit, but later, and follow it within 1% accuracy over 3 decades in time ($t \geqslant 10^5\sigma v^{-1}$). At $R/\sigma = 0.5$, $\gamma(t)$ approximately resembles the curve at $R\sigma = 0.9$ up to $t \approx 10^5\sigma v^{-1}$, but then appears to increase further; this part of the data for $\gamma(t)$, however, suffers from statistical noise and is still far from $2/z_{\mathrm{lat}}$. On basis of the critical MSDs, we conclude that the magnetic value of the dynamic exponent, $z \approx 3.44$, is (within our accuracy) independent of $R$, and $z$ is clearly different from $z_{\mathrm{lat}}$. Thus, the critical dynamics due to field-induced localization is not described by the lattice universality class.

Finally, we note that the discrepancy in the dynamic exponents is unlikely to arise from asymptotic corrections to scaling.

The standard scenario predicts that [5]

$$\delta r^2(t \to \infty) \simeq A t^{2/z}(1 + C t^{-y}) \qquad \text{(S6)}$$

with universal correction exponent $y = \Omega d_\mathrm{f}/d_\mathrm{w} \approx 0.521$ [4], using $\Omega = 72/91$ [6] and non-universal correction amplitude $C < 0$. In particular, the critical long-time asymptote is expected to be approached from *below*, at variance to what we observe for the magnetic transition. Even if a positive sign of $C$ is admitted, Fig. S2 demonstrates impressively that deviations of the MSD from $At^{2/z_\mathrm{lat}}$ cannot be rationalized by the (lattice) exponent $y$.

### D. Consistency checks

The question remains whether the exponents $z$ and $\mu$ are independent of the magnetic field (Fig. S2) or not (Fig. S1). Moreover, both exponents are linked via Eq. (S4). The analysis of $\gamma(t)$ appears to be the most sensitive tool, and the results in Table S1 motivate the hypothesis of a common value $\gamma = 0.582 \pm 0.010$, or equivalently $z = 3.44 \pm 0.06$ and $\mu = 1.82 \pm 0.08$, at all magnetic field parameters $1/R$, which needs to be tested against the data for the diffusion constant. (The increased uncertainties reflect also errors not covered by the fitting procedure.) Figure 3b of the main paper shows an alternative rectification plot of $D$ with conductivity exponent $\mu = 1.82$. Considering only data points close to the transition, the data are indeed compatible with such a scenario. The figure suggests that the asymptotic regime, however, sets in earlier (at larger $\varepsilon$) for stronger fields.

## II. FREQUENCY-DEPENDENT CONDUCTIVITY

From the simulated MSDs we also make a prediction for the complex a.c. conductivity $\Sigma(\omega)$ of the sample, which can be measured in experiments. The connection is based on the Einstein–Kubo relation $\Sigma(\omega) \propto Z(\omega)$ [7], where $Z(\omega)$ is the Fourier–Laplace transform of the velocity auto-correlation function:

$$Z(\omega) = \lim_{\varepsilon \downarrow 0} \frac{1}{d} \int_0^\infty \langle \mathbf{v}(t) \cdot \mathbf{v}(0) \rangle \, \mathrm{e}^{\mathrm{i}(\omega + \mathrm{i}\varepsilon)t} \mathrm{d}t \,. \qquad \text{(S7)}$$

After some manipulation, one finds

$$Z(\omega) = D - \mathrm{i}\omega \int_0^\infty \left[ \frac{1}{2d} \frac{\mathrm{d}}{\mathrm{d}t} \delta r^2(t) - D \right] \mathrm{e}^{\mathrm{i}\omega t} \, \mathrm{d}t \,, \qquad \text{(S8)}$$

which is conveniently evaluated numerically from the simulation data for $\delta r^2(t)$ and $D$ using the simplified Filon algorithm, see ref. [8] for details. Figure S3 shows the results for different densities at fixed magnetic field.

In the low-frequency limit, $\Sigma(\omega \to 0) = \Sigma_0 \propto D$, one recovers the d.c. conductivity $\Sigma_0$, corresponding to the long-time diffusion constant $D$. Our data (Fig. S3) exhibit a suppression of $\Sigma_0(n^*)$ by almost 3 orders of magnitude as the magnetic transition is approached, $n^* \to n^*_m$. At high frequencies, $\Sigma(\omega)$ is almost insensitive to changes of the obstacle density. Thus, a pronounced frequency-dependence develops in $\Sigma(\omega)$, which is most prominently manifested in an anomalous power-law dispersion right at the transition:

$$\Sigma(\omega \to 0; n^* = n^*_m) \sim \omega^{1-2/z} \,. \qquad \text{(S9)}$$

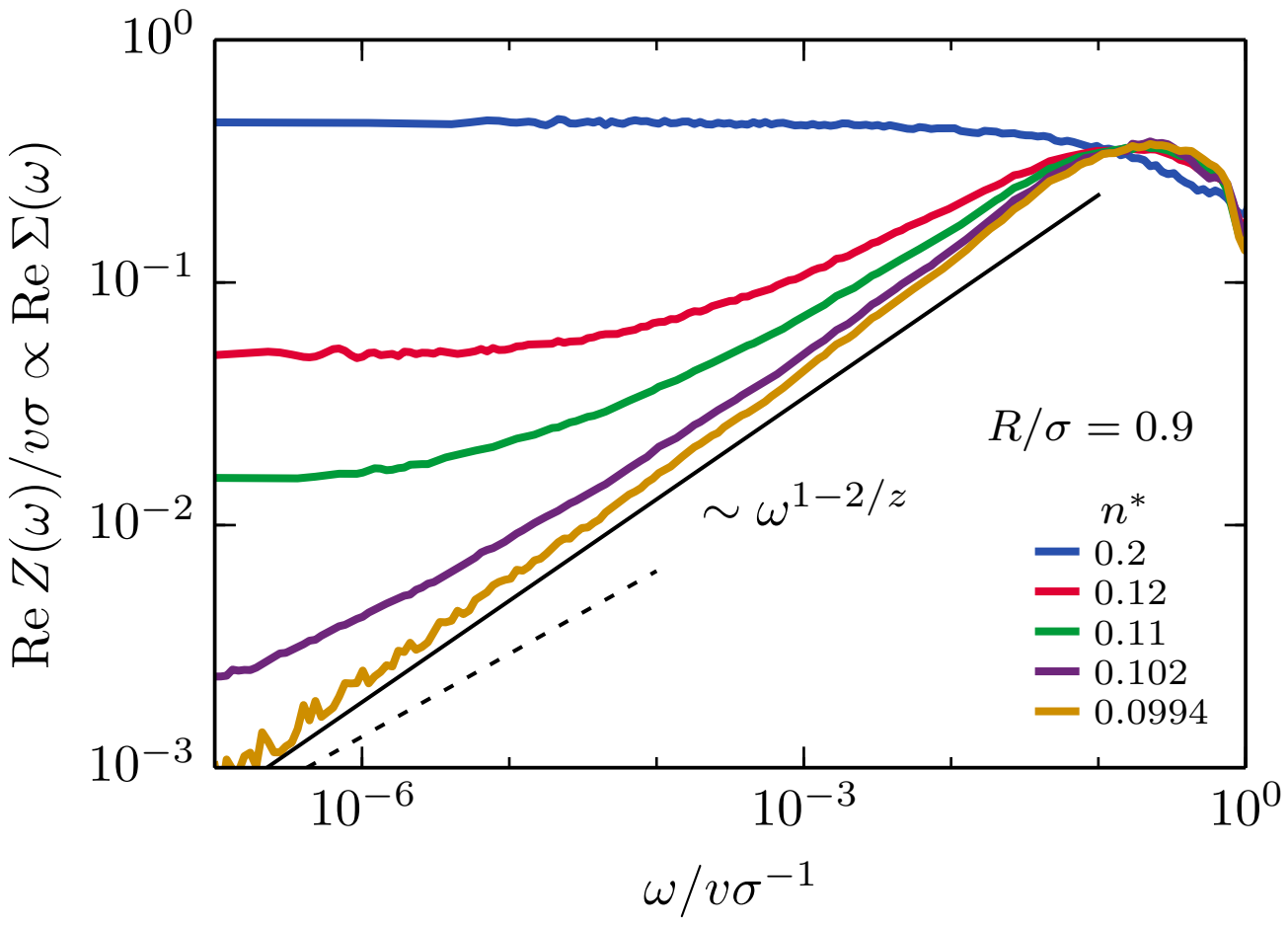

Figure S3. Loss part of the frequency-dependent conductivity $\mathrm{Re}\, \Sigma(\omega) \propto \mathrm{Re}\, Z(\omega)$ for obstacle densities approaching the field-induced localization transition at fixed magnetic field ($R = 0.9\sigma$). The data were obtained from a numerical Fourier–Laplace transformation of the mean-square displacements, see text. The solid straight line indicates the critical asymptote, $\Sigma(\omega \to 0) \sim \omega^{1-2/z}$ with $z = 3.44$, the dashed straight line shows $\omega^{1-2/z_\mathrm{lat}}$ for comparison.

This power law and the slow asymptotic approach to it are both inherited from the MSD, and both are corroborated nicely by our data. It is also evident that the anomalous dispersion $\Sigma(\omega) \sim \omega^{1-2/z_\mathrm{lat}}$, as expected from the lattice universality class, does not describe the data.

### Appendix A: Power-law regression

The exponents in Figs. S1 and S2 were determined from the regression of the power law model,

$$y(x) = \alpha (x/x_0)^\beta \,, \qquad \text{(S1)}$$

using a weighted least-squares approach. Take $n$ observations $y = (y_1, \ldots, y_n)$ with uncertainties $(\Delta y_1, \ldots, \Delta y_n)$ at given input $x = (x_1, \ldots, x_n)$. First, the data are transformed to their logarithms, $\xi_i = \log(x_i)$ and $\eta_i = \log(y_i)$, and the uncertainties $\Delta y_i$ are converted to $\sigma_i = \log(1 + \Delta y_i / y_i)$, employing the geometric coefficient of variation of the log-normal distribution. Let us define a weighted average and a weighted 2-norm based on $(\sigma_1, \ldots, \sigma_n)$:

$$\bar{z} := \sum_{i=1}^{n} w_i z_i \,, \quad \|z\|^2 := \sum_{i=1}^{n} w_i z_i^2 \,, \quad w_i := \frac{\sigma_i^{-2}}{\sum_i \sigma_i^{-2}} \,.$$

Then, minimization of the weighted residual

$$\left\| \eta - \hat{\alpha} + \hat{\beta}(\xi_i - \bar{\xi}) \right\|^2$$

yields estimates $\hat{\alpha}, \hat{\beta}$ for the model parameters; further, $x_0 := \exp(\bar{\xi})$. From the theory of linear regression, the uncertainties on $\hat{\alpha}$ and $\hat{\beta}$ are given by

$$(\Delta\alpha)^2 = 1 \Big/ \sum_i \sigma_i^{-2}, \quad (\Delta\beta)^2 = \sum_i \sigma_i^{-2} \Big/ \left\| \xi - \bar{\xi} \right\|^2.$$

It is convenient to let an ordinary least-squares routine operate on the transformed data $\tilde{\xi}_i = \xi_i / \sigma_i$ and $\tilde{\eta}_i = \eta_i / \sigma_i$.

For mere amplitude fits in regression plots (e.g., Figs. 3a,b of the main paper), the exponent is fixed to $\beta = 0$. In these cases, we have used weighted averages of the original data with $w_i \propto (\Delta y_i)^{-2}$:

$$\hat{\alpha} = \bar{y}\,, \qquad (\Delta\alpha)^2 = \frac{\| y - \hat{\alpha} \|}{n-1}\,. \tag{S2}$$